\documentclass[cameraready]{Interspeech}

\usepackage{graphicx}
\usepackage{subcaption}
\usepackage{booktabs} 

\usepackage{hyperref}
\usepackage{amsmath}
\usepackage{amssymb}
\usepackage{mathtools}
\usepackage{amsthm}

\usepackage{multirow}
\usepackage[table]{xcolor}
\usepackage{xcolor}
\definecolor{darkpurple}{RGB}{120,60,160}

\usepackage{lipsum}
\usepackage{listings}
\usepackage{verbatim}

\usepackage[capitalize,noabbrev]{cleveref}

\usepackage[
backend=biber,
style=ieee,
maxbibnames=5,
maxcitenames=3,
doi=false,isbn=false,url=false,eprint=false, date=year,url=false
]{biblatex} 
\defbibheading{bibliography}[\refname]{}

\newcommand{\symbolvar}[1]{\boldsymbol{\mathrm{#1}}}
\newcommand{\symbolvec}[1]{\boldsymbol{#1}}

\newcommand{\symboldomain}[1]{\mathbb{#1}}
\newcommand{\symbolset}[1]{\mathcal{#1}}
\newcommand{\symbolloss}[1]{\mathcal{#1}}
\newcommand{\symboltag}[1]{\texttt{#1}}

\theoremstyle{plain}

\theoremstyle{definition}

\theoremstyle{remark}

\newcommand{\testdf}{{\texttt{DFE24}}}
\newcommand{\testadd}{{\texttt{ADD23}}}
\newcommand{\testfor}{{\texttt{FoR}}}
\newcommand{\testdv}{{\texttt{DV}}}
\newcommand{\testitw}{{\texttt{ItW}}}

\newcounter{rownumid}
\newcommand{\rowid}{\stepcounter{rownumid}\arabic{rownumid}}

\newcommand{\algogrpo}{{GRPO}}
\newcommand{\algogrpos}{$\text{GRPO}_{\text{s}}$}
\usepackage[textsize=tiny]{todonotes}

\title{Does Fine-tuning by Reinforcement Learning Improve Generalization in Binary Speech Deepfake Detection?}

\author[affiliation={1}]{Xin}{Wang}
\author[affiliation={1}]{Wanying}{Ge}
\author[affiliation={1}]{Junichi}{Yamagishi}

\address{
    $^1$ National Institute of Informatics, Japan
}
\email{wangxin@nii.ac.jp}


\begin{document}
\maketitle

\begin{abstract}
Building speech deepfake detection models that are generalizable to unseen attacks remains a challenging problem. Although the field has shifted toward a pre-training and fine-tuning paradigm using speech foundation models, most approaches rely solely on supervised fine-tuning (SFT). Inspired by the field of large language models, wherein reinforcement learning (RL) is used for model fine-tuning, we investigate the impact of RL, specifically Group Relative Policy Optimization (GRPO). The results from experiments using multiple detectors and test sets indicate that pure GRPO-based fine-tuning improves performance on out-of-domain test sets while maintaining performance on target-domain test data. This approach outperforms both SFT-only and hybrid setups. Our ablation studies further suggest that the negative reward in GRPO may be a key factor in this improvement.
\end{abstract}

\section{Introduction}
\label{sec:intro}
\label{sec:training_paradigm}
Speech deepfake detection is an important, if not the ultimate, countermeasure against speech deepfake~\cite{eureport2024}. It is by convention formulated as a binary classification task---given an input utterance, a speech deepfake detection model makes a decision on whether the input is uttered by a real human speaker~\cite{wu2017asvspoof}. The goal is to develop a detection model that is generalizable to unseen deepfake types, data from unseen speakers, etc.  

\begin{figure}
    \centering
    \includegraphics[width=\columnwidth, trim=15 390 15 5, clip]{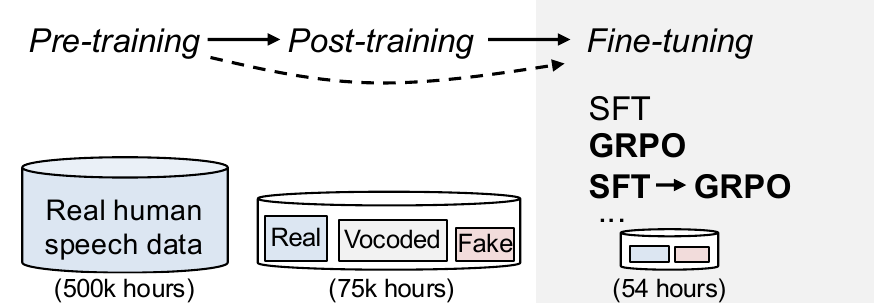}
    \caption{Illustration of multi-stage training pipeline for speech deepfake detection models. This paper investigates different fine-tuning algorithms in shaded area.}
    \label{fig:framework}
\end{figure}

Training the detection model using a carefully curated dataset, which contains real and fake data, has been a common practice over the past decade~\cite{todiscoASVspoof2019Future2019, yiADD2023}. A surging trend in the field, however, is the framework of pre-training plus fine-tuning~\cite{martin-donas_vicomtech_2022,tak_automatic_2022,zhangAudio2024}. The pre-training stage~\cite{mohamed_self-supervised_2022}, usually conducted by teams with abundant computation power and data, uses a huge amount of diverse real data and self-supervised learning (SSL) criteria to build speech foundation models (e.g., wav2vec 2.0~\cite{wav2vec2}). The foundation model is then tuned using supervised fine-tuning (SFT) on small data for speech deepfake detection. Models built from this two-stage approach have demonstrated top performance in recent benchmarks~\cite{wang2024asvspoof5,dowerah_speech_2026}. 
The pre-training and fine-tuning framework was recently extended to include a post-training phase in the middle~\cite{antideepfake_2025}, which is illustrated in Figure~\ref{fig:framework}. The post-training leverages a large amount of real, fake, and simulated fake data (e.g., vocoded data~\cite{antideepfake_2025}) from various domains to adapt the general-purpose foundation models, assuming that the adapted model can produce representations more suitable for deepfake detection, source tracing~\cite{klein_source_2024}, and other related downstream tasks.

It is obvious that the above multi-stage training pipeline resembles that of large language models (LLMs)~\cite{guo_deepseek-r1_2025}. 
An LLM also undergoes pre-training to learn general language patterns, optionally mid-training~\cite{mo_mid-training_2025}, followed by fine-tuning to align the LLM with human preferences or specific tasks. 
An issue common to both fields, however, is that the model fine-tuned towards a specific task or data domain is prone to degradation in other tasks or data domains, which is also known as catastrophic forgetting~\cite{wangComprehensive2024}.
The LLM field has explored various fine-tuning setups, for example, combining SFT and 
reinforcement learning (RL)~\cite{guo_deepseek-r1_2025}, or RL-only fine-tuning via the Group Relative Policy Optimization (GRPO)~\cite{guo_deepseek-r1_2025}. Regardless of the pipeline, RL was found to be an essential step into improving the LLM performance across tasks and domains~\cite{guo_deepseek-r1_2025} and better preserving the prior knowledge acquired during pre-training~\cite{shenfeld_rls_2025}. 

Because most studies on speech deepfake detection only conducted SFT (even for post-training~\cite{antideepfake_2025}), 
we were inspired by the progress in the LLM field and investigated \emph{whether RL-based fine-tuning can improve the generalization performance of speech deepfake detectors}. 
We thus applied RL, particularly GRPO, to fine-tune speech deepfake detection models, as illustrated in Figure~\ref{fig:framework}. Our findings and contributions are summarized below.
\begin{itemize}
    \item We applied GRPO to multiple post-trained SSL-based speech deepfake detectors and found that pure GRPO-based fine-tuning improved the detection performance on all the out-of-domain test sets, outperforming other fine-tuning setups such as only SFT or SFT plus GRPO. 
    \item We conducted ablation studies and found that the negative reward in GRPO may be a key component for this improvement. The regularization term in GRPO was less impactful.  
    \item We measured the drift of the test data distributions using a recently published tool~\cite{wang_towards_2026} and found evidence to support the above findings.
\end{itemize}
A recent study applied GRPO to LLM-based speech deepfake detectors~\cite{xie_interpretable_2026}, but our above findings are new, to the best of our knowledge, to the mainstream approach using SSL-based front ends. We will release the code after the paper review. 

\section{Methods}
\label{sec:method}
\subsection{Training paradigm and scope of this paper}
Given an input waveform $\symbolvec{x}\in\symboldomain{R}^T$, our detection model is expected to produce a label $y\in\symbolset{Y}=\{\symboltag{REAL}, \symboltag{FAKE}\}$ that flags whether the input is real or not. 
For model training, we tune the parameter set $\Theta$ of a model ${p}_\Theta(\symbolvar{y} | \symbolvar{\symbolvec{x}})$ via a training pipeline.
When using the trained model for making decisions, we compute the probability ${p}_\Theta(\symbolvar{y}=\symboltag{REAL} | \symbolvar{\symbolvec{x}}=\symbolvec{x}^{\text{test}})$ (or its logit) for an input sample $\symbolvec{x}^{\text{test}}$. If the probability is higher than a pre-defined threshold, we decide $\widehat{y}=\symboltag{REAL}$; otherwise, $\widehat{y}=\symboltag{FAKE}$.

In this paper, we consider the multi-stage training pipeline (Figure~\ref{fig:framework}) and detection models with an SSL-based front end and a binary classification back end.  
We assume that the parameter set $\Theta$, including both the front and back ends, has been post-trained, and we only focus on fine-tuning $\Theta$ using a relatively small-scale training set $\symbolset{D}=\{\symbolvec{x}^{(k)}, y^{(k)} \}_k$ from a specific domain.\footnote{
To avoid notion clutter, we drop the sample index $(k)$ unless it is necessary. We do not distinguish between random variables and their values (e.g., $\symbolvar{y}$ vs. $y$) unless the distinction is necessary.}
Note that, for a conventional setup with only pre-training and fine-tuning, $\Theta$ consists of the parameters of the pre-trained front end and a randomly initialized back end.

\subsection{Supervised fine-tuning}
The most common way of fine-tuning $\Theta$ is using SFT. It minimizes the negative log-likelihood $\symbolloss{L}_{\text{ne}}$ of the training data $\symbolset{D}$, which can be written as
\begin{align}
\symbolloss{L}_{\text{ne}}(\Theta) &= -\symboldomain{E}_{\symbolvec{x}, y\sim\symbolset{D}}\Big[\log p_\Theta(y | \symbolvec{x})\Big] \label{eq:sft} \\
&\approx-\sum_{i=1}^{|\symbolset{D}|}\sum_{\tilde{y}\in\symbolset{Y}}\delta(\tilde{y}=y^{(i)})\log p_{\Theta}(\tilde{y}|\symbolvec{x}^{(i)}),\label{eq:sft2} 
\end{align}
where $\delta(\cdot)$ is an indicator function. 

\subsection{Reinforcement learning}
\label{sec:method:rl}
RL is an alternative to SFT. Among many RL-based algorithms,  we consider the policy-based algorithm that defines the loss as
\begin{align}
    \symbolloss{L}_{\text{RL}}(\Theta) &= -\symboldomain{E}_{\symbolvec{x}\sim\symbolset{D}, \tilde{y}\sim p_{\widehat{\Theta}}(\symbolvar{y} | \symbolvar{x})}\Big[\symbolloss{A}(\tilde{y},y)\frac{p_\Theta(\tilde{y} | \symbolvec{x})}{p_{\widehat{\Theta}}(\tilde{y} | \symbolvec{x})}-\beta\symbolloss{L}_{\text{p}}({\dot{\Theta}}, {{\Theta}})\Big],\label{eq:rl}
\end{align}
where $\tilde{y}$ is a label sampled from the model with a snapshot of the `old' parameter set $\widehat{\Theta}$ (e.g., the one from the previous training epoch). An advantage function 
$\symbolloss{A}(\tilde{y},y): \symbolset{Y}\times\symbolset{Y}\rightarrow\symboldomain{R}$ measures how well the sampled label $\tilde{y}$ matches the ground truth $y$. 
The regularization term $\symbolloss{L}_{\text{p}}$ penalizes the distance between the fine-tuned model and a reference model (e.g., a pre-/post-trained one with a frozen parameter set $\dot{\Theta}$), even though the regularization is argued to be non-essential~\cite{shenfeld_rls_2025}. It can be implemented using an approximated Kullback–Leibler divergence~\cite[\S~2.3]{amini_better_2025}.

GRPO is a common RL algorithm for fine-tuning LLMs~\cite{shenfeld_rls_2025}. It enhances Eq.~(\ref{eq:rl}) by sampling multiple instances $\{\tilde{y}_j\}_{j=1}^{G}$ for each training input $\symbolvec{x}$ and computing a group-normalized advantage $\symbolloss{A}(\tilde{y}_i, \{\tilde{y}_j\}_{j=1}^{G}, y)$ using
\begin{align}
\symbolloss{A}(\tilde{y}_i, \{\tilde{y}_j\}_{j=1}^{G}, y) = \frac{r(\tilde{y}_i, y) - \bar{r}(\{\tilde{y}\}_{j=1}^G, y)}{\sigma(\{\tilde{y}\}_{j=1}^G, y)},
\end{align}
where $r(\tilde{y}_i, y)$ is a reward function that measures the goodness of $\tilde{y}_i$ when compared against $y$ or pre-defined rules~\cite{shaoDeepSeekMath2024}, and where $\bar{r}$ and $\sigma$ are the mean and standard deviation of $r$ over the group of sample outputs $\{\tilde{y}_j\}_{j=1}^G$. 

GRPO also clips the ratio $\frac{p_\Theta(\tilde{y} | \symbolvec{x})}{p_{\widehat{\Theta}}(\tilde{y} | \symbolvec{x})}$, which is skipped to avoid notation clutter. 
GRPO also requires three parameter sets: the one to be updated ($\Theta$) at every training step, the `old' set ($\hat{\Theta}$) that is updated less frequently, and the frozen reference ($\dot{\Theta}$). For simplification, we can set $\hat{\Theta}=\Theta$ and add a stop-gradient operator in $\frac{p_\Theta(\tilde{y} | \symbolvec{x})}{[p_{{\Theta}}(\tilde{y} | \symbolvec{x})]_\text{sg}}$ in Eq.~(\ref{eq:rl}). We can further remove the clipping function. This \emph{simplified} GRPO (\algogrpos) updates the model after every batch of sampled data, similar to on-policy RL by definition~\cite{rlhf2024}.

\subsection{Applying GRPO to deepfake detector}
\label{sec:grpo:setup}
Fine-tuning using GRPO can be directly applied to binary deepfake detectors. The essential difference from the case of LLM is that the sampled output $\tilde{y}$ for the deepfake detector is either $\symboltag{FAKE}$ or $\symboltag{REAL}$, rather than a sequence of tokens. The reward can be simply defined as an indicator function $r(\tilde{y}_i, y)=\delta(\tilde{y}_i=y)$ so that $r=1$ for $\tilde{y}_i =y$; otherwise, $0$. 
The reference parameter $\dot{\Theta}$ can be that of the post-trained model. 

In addition to directly applying GRPO, we intend to understand the differences between GRPO and SFT that potentially affect the generalization capability of the fine-tuned detection model. Notable differences are listed below.
\begin{itemize}
    \item[a)] \textbf{Regularization term} $\symbolloss{L}_{\text{p}}$: Regularization w.r.t. a `good old' model is assumed to alleviate the issue of catastrophic forgetting during fine-tuning~\cite{wang_comprehensive_2024}. Although a few studies argue that $\symbolloss{L}_{\text{p}}$ is not essential for LLMs~\cite{yu_dapo_2025}, it may affect the performance of the binary detector, the output of which is binary. 
    \item[b)] \textbf{Advantage function} $\symbolloss{A}(\tilde{y}_i, \{\tilde{y}_j\}_{j=1}^{G}, y)$: the advantage function in GRPO accumulates positive and negative rewards from \emph{both} matched ($\tilde{y}_i=y$) and unmatched ($\tilde{y}_i\neq y$) samples, even if we define a 1-0 reward $r(\tilde{y}_i, y)=\delta(\tilde{y}_i=y)$.\footnote{Specifically, we obtain $\symbolloss{A}(\tilde{y}_i, \{\tilde{y}_j\}_{j=1}^{G}, y)= (1-\bar{r})/\sigma$ and $-\bar{r}/\sigma$ for matched ($r(\tilde{y}_i, y)=1$) and unmatched ($r(\tilde{y}_i, y)=0$) samples.} This is different from SFT where the loss is computed using only the ground-truth, i.e., $\widehat{y}$ is always equal to $y$ in Eq.~(\ref{eq:sft2}). 
\end{itemize}
To analyze the impact of the two factors, we introduce a few variants and compare their performance through experiments:
\begin{itemize}
    \item \algogrpo{} ($\beta=0$): this variant sets the weight of $\symbolloss{L}_{\text{p}}$ to 0 and testifies whether GRPO benefits from the regularization;
    \item \algogrpo{} using a large $\beta$ (GRPO ($\beta=1$))\footnote{$\beta=1$ is 25 times larger than the default $\beta=0.04$~\cite{shaoDeepSeekMath2024}.}: it examines whether excess regularization degrades performance. 
    \item \algogrpo{} without any negative sample (\algogrpo{} w/o neg.): it defines $\symbolloss{A}(\tilde{y}_i, y)=r(\tilde{y}_i, y)=\delta(\tilde{y}_i=y)$, hence removes the negative case $\widehat{y}_i\neq y$. This variant is similar to SFT plus the regularization term (ref. supplementary material).
\end{itemize}
The above variants can be applied to the simplified GRPO (\algogrpos), but we skip these configurations in this paper.

\begin{table*}[t!]
    \centering
    \setlength{\tabcolsep}{3.5pt}
    \caption{Test set EERs (\%). Each EER was average after \emph{three} training-evaluation rounds. Columns `ave.' list average EERs of in- or out-of-domain. GRPO by default uses $\beta=0.04$ (\S\ref{sec:exp:config}). In each column, higher EER is in darker shade. }
    \vspace{-2mm}
    \setcounter{rownum}{0} 
    \begin{tabular}{llllrrrrrrrrrrr|r}
\toprule
 & \multirow{3}{*}{\shortstack{Pre-\\train}} 
 & \multirow{3}{*}{\shortstack{Post-\\train}} 
 & \multirow{3}{*}{\shortstack{Fine-\\tune}} & 
 \multicolumn{6}{c}{\testdf{} eva. set (in-domain)} & \multicolumn{5}{c}{Other eva. sets (out-of-domain)} & \multirow{3}{*}{\shortstack{Row \\ ID}} \\ \cmidrule(lr){5-10}\cmidrule(lr){11-15}
    & & & &  4 s   &  10 s  &  13 s  &  30 s  &  50 s & ave. & \testadd &  \testfor  &  \testdv   &  \testitw & ave. & \\ 
\midrule
\multirow{13}{*}{\rotatebox{90}{XLS-R-2B}}
      & $\checkmark$ & $\checkmark$ & $\times$ 
      & \cellcolor[rgb]{0.70, 0.70, 0.70} 27.73 & \cellcolor[rgb]{0.77, 0.77, 0.77} 23.45 & \cellcolor[rgb]{0.79, 0.79, 0.79} 21.74 & \cellcolor[rgb]{0.81, 0.81, 0.81} 20.62 & \cellcolor[rgb]{0.82, 0.82, 0.82} 19.64 & \cellcolor[rgb]{0.78, 0.78, 0.78} 22.64 & \cellcolor[rgb]{0.97, 0.97, 0.97} 4.67 & \cellcolor[rgb]{0.99, 0.99, 0.99} 2.61 & \cellcolor[rgb]{0.99, 0.99, 0.99} 2.23 & \cellcolor[rgb]{1.00, 1.00, 1.00} 1.24 & \cellcolor[rgb]{0.99, 0.99, 0.99} 2.69 & \rowid\\ 
    \cmidrule{2-15}
   & \multirow{8}{*}{$\checkmark$} & \multirow{8}{*}{$\checkmark$} & SFT
   & \cellcolor[rgb]{0.91, 0.91, 0.91} 12.17 & \cellcolor[rgb]{0.93, 0.93, 0.93} 10.28 & \cellcolor[rgb]{0.93, 0.93, 0.93} 10.11 & \cellcolor[rgb]{0.94, 0.94, 0.94} 9.50 & \cellcolor[rgb]{0.94, 0.94, 0.94} 9.23 & \cellcolor[rgb]{0.93, 0.93, 0.93} 10.26 & \cellcolor[rgb]{0.96, 0.96, 0.96} 6.09 & \cellcolor[rgb]{0.98, 0.98, 0.98} 3.92 & \cellcolor[rgb]{0.94, 0.94, 0.94} 8.75 & \cellcolor[rgb]{0.96, 0.96, 0.96} 6.35 & \cellcolor[rgb]{0.96, 0.96, 0.96} 6.28 &\rowid\\ 
   &  &  & \algogrpos
   & \cellcolor[rgb]{0.90, 0.90, 0.90} 12.45 & \cellcolor[rgb]{0.92, 0.92, 0.92} 11.07 & \cellcolor[rgb]{0.92, 0.92, 0.92} 11.06 & \cellcolor[rgb]{0.92, 0.92, 0.92} 10.44 & \cellcolor[rgb]{0.92, 0.92, 0.92} 10.98 & \cellcolor[rgb]{0.92, 0.92, 0.92} 11.20 & \cellcolor[rgb]{0.98, 0.98, 0.98} 3.95 & \cellcolor[rgb]{1.00, 1.00, 1.00} 0.65 & \cellcolor[rgb]{0.96, 0.96, 0.96} 6.72 & \cellcolor[rgb]{0.99, 0.99, 0.99} 2.44 & \cellcolor[rgb]{0.98, 0.98, 0.98} 3.44 &\rowid\\ 
   &  &  & \algogrpo
   & \cellcolor[rgb]{0.92, 0.92, 0.92} 11.06 & \cellcolor[rgb]{0.93, 0.93, 0.93} 9.99 & \cellcolor[rgb]{0.93, 0.93, 0.93} 9.90 & \cellcolor[rgb]{0.94, 0.94, 0.94} 9.45 & \cellcolor[rgb]{0.94, 0.94, 0.94} 9.27 & \cellcolor[rgb]{0.93, 0.93, 0.93} 9.93 & \cellcolor[rgb]{0.97, 0.97, 0.97} 5.34 & \cellcolor[rgb]{1.00, 1.00, 1.00} 0.47 & \cellcolor[rgb]{0.99, 0.99, 0.99} 2.76 & \cellcolor[rgb]{0.99, 0.99, 0.99} 2.19 & \cellcolor[rgb]{0.99, 0.99, 0.99} 2.69 &\rowid\\ 
   \cmidrule{4-15}
   &  &  & SFT $\rightarrow$ \algogrpos   
   & \cellcolor[rgb]{0.91, 0.91, 0.91} 12.17 & \cellcolor[rgb]{0.93, 0.93, 0.93} 10.20 & \cellcolor[rgb]{0.93, 0.93, 0.93} 10.07 & \cellcolor[rgb]{0.94, 0.94, 0.94} 9.36 & \cellcolor[rgb]{0.94, 0.94, 0.94} 9.43 & \cellcolor[rgb]{0.93, 0.93, 0.93} 10.24 & \cellcolor[rgb]{0.96, 0.96, 0.96} 6.42 & \cellcolor[rgb]{0.98, 0.98, 0.98} 3.11 & \cellcolor[rgb]{0.95, 0.95, 0.95} 8.20 & \cellcolor[rgb]{0.96, 0.96, 0.96} 6.24 & \cellcolor[rgb]{0.96, 0.96, 0.96} 5.99 &\rowid\\ 
   &  &  & SFT $\rightarrow$ \algogrpo   
   & \cellcolor[rgb]{0.91, 0.91, 0.91} 11.53 & \cellcolor[rgb]{0.93, 0.93, 0.93} 9.94 & \cellcolor[rgb]{0.93, 0.93, 0.93} 9.88 & \cellcolor[rgb]{0.94, 0.94, 0.94} 8.93 & \cellcolor[rgb]{0.94, 0.94, 0.94} 8.76 & \cellcolor[rgb]{0.93, 0.93, 0.93} 9.81 & \cellcolor[rgb]{0.96, 0.96, 0.96} 5.78 & \cellcolor[rgb]{0.99, 0.99, 0.99} 2.75 & \cellcolor[rgb]{0.95, 0.95, 0.95} 7.04 & \cellcolor[rgb]{0.96, 0.96, 0.96} 5.89 & \cellcolor[rgb]{0.97, 0.97, 0.97} 5.37 &\rowid\\ 
   \cmidrule{4-15}
   &  &  & \algogrpo{} ($\beta=0$) 
   & \cellcolor[rgb]{0.91, 0.91, 0.91} 11.58 & \cellcolor[rgb]{0.93, 0.93, 0.93} 10.32 & \cellcolor[rgb]{0.93, 0.93, 0.93} 10.00 & \cellcolor[rgb]{0.93, 0.93, 0.93} 9.64 & \cellcolor[rgb]{0.94, 0.94, 0.94} 9.51 & \cellcolor[rgb]{0.93, 0.93, 0.93} 10.21 & \cellcolor[rgb]{0.97, 0.97, 0.97} 4.43 & \cellcolor[rgb]{1.00, 1.00, 1.00} 0.47 & \cellcolor[rgb]{0.96, 0.96, 0.96} 6.50 & \cellcolor[rgb]{0.99, 0.99, 0.99} 2.71 & \cellcolor[rgb]{0.98, 0.98, 0.98} 3.53 &\rowid\\ 
   &  &  & \algogrpo{} ($\beta=1$) 
   & \cellcolor[rgb]{0.82, 0.82, 0.82} 19.37 & \cellcolor[rgb]{0.87, 0.87, 0.87} 15.61 & \cellcolor[rgb]{0.87, 0.87, 0.87} 15.27 & \cellcolor[rgb]{0.88, 0.88, 0.88} 14.63 & \cellcolor[rgb]{0.88, 0.88, 0.88} 14.24 & \cellcolor[rgb]{0.86, 0.86, 0.86} 15.83 & \cellcolor[rgb]{0.98, 0.98, 0.98} 4.02 & \cellcolor[rgb]{0.99, 0.99, 0.99} 1.34 & \cellcolor[rgb]{0.99, 0.99, 0.99} 1.66 & \cellcolor[rgb]{1.00, 1.00, 1.00} 1.13 & \cellcolor[rgb]{0.99, 0.99, 0.99} 2.04 &\rowid\\ 
   &  &  & GRPO w/o neg. & 
   \cellcolor[rgb]{0.89, 0.89, 0.89} 13.30 & \cellcolor[rgb]{0.91, 0.91, 0.91} 11.60 & \cellcolor[rgb]{0.92, 0.92, 0.92} 11.23 & \cellcolor[rgb]{0.92, 0.92, 0.92} 10.64 & \cellcolor[rgb]{0.93, 0.93, 0.93} 10.31 & \cellcolor[rgb]{0.91, 0.91, 0.91} 11.41 & \cellcolor[rgb]{0.96, 0.96, 0.96} 6.69 & \cellcolor[rgb]{1.00, 1.00, 1.00} 0.83 & \cellcolor[rgb]{0.98, 0.98, 0.98} 3.09 & \cellcolor[rgb]{0.98, 0.98, 0.98} 3.14 & \cellcolor[rgb]{0.98, 0.98, 0.98} 3.44 &\rowid\\
   \cmidrule{2-15}
      & \multirow{2}{*}{$\checkmark$} & \multirow{2}{*}{$\times$} & SFT 
      & \cellcolor[rgb]{0.90, 0.90, 0.90} 12.24 & \cellcolor[rgb]{0.93, 0.93, 0.93} 10.34 & \cellcolor[rgb]{0.93, 0.93, 0.93} 10.20 & \cellcolor[rgb]{0.93, 0.93, 0.93} 9.63 & \cellcolor[rgb]{0.93, 0.93, 0.93} 9.83 & \cellcolor[rgb]{0.92, 0.92, 0.92} 10.45 & \cellcolor[rgb]{0.84, 0.84, 0.84} 17.81 & \cellcolor[rgb]{0.66, 0.66, 0.66} 29.83 & \cellcolor[rgb]{0.86, 0.86, 0.86} 16.21 & \cellcolor[rgb]{0.89, 0.89, 0.89} 13.49 & \cellcolor[rgb]{0.82, 0.82, 0.82} 19.34 &\rowid\\ 
      &  &  & \algogrpos & \cellcolor[rgb]{0.88, 0.88, 0.88} 14.33 & \cellcolor[rgb]{0.90, 0.90, 0.90} 12.65 & \cellcolor[rgb]{0.91, 0.91, 0.91} 12.02 & \cellcolor[rgb]{0.92, 0.92, 0.92} 11.27 & \cellcolor[rgb]{0.92, 0.92, 0.92} 10.87 & \cellcolor[rgb]{0.90, 0.90, 0.90} 12.23 & \cellcolor[rgb]{0.77, 0.77, 0.77} 22.88 & \cellcolor[rgb]{0.66, 0.66, 0.66} 29.71 & \cellcolor[rgb]{0.76, 0.76, 0.76} 23.67 & \cellcolor[rgb]{0.77, 0.77, 0.77} 23.03 & \cellcolor[rgb]{0.75, 0.75, 0.75} 24.82 &\rowid\\ 
    &  &  & \algogrpo 
      & \cellcolor[rgb]{0.90, 0.90, 0.90} 12.93 & \cellcolor[rgb]{0.92, 0.92, 0.92} 11.18 & \cellcolor[rgb]{0.92, 0.92, 0.92} 11.06 & \cellcolor[rgb]{0.93, 0.93, 0.93} 10.24 & \cellcolor[rgb]{0.93, 0.93, 0.93} 10.35 & \cellcolor[rgb]{0.92, 0.92, 0.92} 11.15 & \cellcolor[rgb]{0.85, 0.85, 0.85} 17.57 & \cellcolor[rgb]{0.74, 0.74, 0.74} 25.68 & \cellcolor[rgb]{0.75, 0.75, 0.75} 25.19 & \cellcolor[rgb]{0.83, 0.83, 0.83} 19.00 & \cellcolor[rgb]{0.79, 0.79, 0.79} 21.86 &\rowid\\ 
   \midrule\midrule
   \multirow{4}{*}{\rotatebox{90}{MMS-1B}} 
   & $\checkmark$ & $\checkmark$ & $\times$ 
      & \cellcolor[rgb]{0.70, 0.70, 0.70} 27.68 & \cellcolor[rgb]{0.77, 0.77, 0.77} 23.49 & \cellcolor[rgb]{0.77, 0.77, 0.77} 22.94 & \cellcolor[rgb]{0.78, 0.78, 0.78} 22.64 & \cellcolor[rgb]{0.78, 0.78, 0.78} 22.59 & \cellcolor[rgb]{0.76, 0.76, 0.76} 23.87 & \cellcolor[rgb]{0.94, 0.94, 0.94} 9.06 & \cellcolor[rgb]{1.00, 1.00, 1.00} 1.23 & \cellcolor[rgb]{0.99, 0.99, 0.99} 2.47 & \cellcolor[rgb]{0.99, 0.99, 0.99} 1.82 & \cellcolor[rgb]{0.98, 0.98, 0.98} 3.64 &\rowid\\  \cmidrule{2-15}
    & \multirow{3}{*}{$\checkmark$} & \multirow{3}{*}{$\checkmark$} & SFT 
      & \cellcolor[rgb]{0.91, 0.91, 0.91} 11.86 & \cellcolor[rgb]{0.93, 0.93, 0.93} 9.87 & \cellcolor[rgb]{0.94, 0.94, 0.94} 9.51 & \cellcolor[rgb]{0.94, 0.94, 0.94} 9.07 & \cellcolor[rgb]{0.94, 0.94, 0.94} 9.08 & \cellcolor[rgb]{0.93, 0.93, 0.93} 9.88 & \cellcolor[rgb]{0.97, 0.97, 0.97} 4.14 & \cellcolor[rgb]{0.99, 0.99, 0.99} 1.35 & \cellcolor[rgb]{0.97, 0.97, 0.97} 5.35 & \cellcolor[rgb]{0.96, 0.96, 0.96} 6.59 & \cellcolor[rgb]{0.97, 0.97, 0.97} 4.36 &\rowid\\  
    &  &  & \algogrpos & \cellcolor[rgb]{0.90, 0.90, 0.90} 12.47 & \cellcolor[rgb]{0.92, 0.92, 0.92} 10.90 & \cellcolor[rgb]{0.93, 0.93, 0.93} 10.22 & \cellcolor[rgb]{0.94, 0.94, 0.94} 9.45 & \cellcolor[rgb]{0.94, 0.94, 0.94} 9.47 & \cellcolor[rgb]{0.92, 0.92, 0.92} 10.50 & \cellcolor[rgb]{0.98, 0.98, 0.98} 3.73 & \cellcolor[rgb]{1.00, 1.00, 1.00} 0.73 & \cellcolor[rgb]{0.99, 0.99, 0.99} 2.68 & \cellcolor[rgb]{0.97, 0.97, 0.97} 4.52 & \cellcolor[rgb]{0.98, 0.98, 0.98} 2.92 &\rowid\\ 
    &  &  & \algogrpo 
      & \cellcolor[rgb]{0.91, 0.91, 0.91} 11.87 & \cellcolor[rgb]{0.92, 0.92, 0.92} 10.41 & \cellcolor[rgb]{0.93, 0.93, 0.93} 9.87 & \cellcolor[rgb]{0.94, 0.94, 0.94} 9.21 & \cellcolor[rgb]{0.94, 0.94, 0.94} 9.47 & \cellcolor[rgb]{0.93, 0.93, 0.93} 10.17 & \cellcolor[rgb]{0.97, 0.97, 0.97} 4.25 & \cellcolor[rgb]{0.99, 0.99, 0.99} 1.78 & \cellcolor[rgb]{0.98, 0.98, 0.98} 2.84 & \cellcolor[rgb]{0.98, 0.98, 0.98} 2.98 & \cellcolor[rgb]{0.98, 0.98, 0.98} 2.96 &\rowid\\ \midrule\midrule
    \multirow{4}{*}{\rotatebox{90}{MMS-300M}} 
   & $\checkmark$ & $\checkmark$ & $\times$ &
      \cellcolor[rgb]{0.61, 0.61, 0.61} 32.87 & \cellcolor[rgb]{0.66, 0.66, 0.66} 30.08 & \cellcolor[rgb]{0.67, 0.67, 0.67} 29.38 & \cellcolor[rgb]{0.72, 0.72, 0.72} 26.33 & \cellcolor[rgb]{0.71, 0.71, 0.71} 27.51 & \cellcolor[rgb]{0.67, 0.67, 0.67} 29.23 & \cellcolor[rgb]{0.95, 0.95, 0.95} 7.96 & \cellcolor[rgb]{0.99, 0.99, 0.99} 1.51 & \cellcolor[rgb]{0.99, 0.99, 0.99} 2.35 & \cellcolor[rgb]{0.98, 0.98, 0.98} 2.91 & \cellcolor[rgb]{0.98, 0.98, 0.98} 3.68 &\rowid\\   \cmidrule{2-15}
    & \multirow{3}{*}{$\checkmark$} & \multirow{3}{*}{$\checkmark$} & SFT 
      & \cellcolor[rgb]{0.88, 0.88, 0.88} 14.45 & \cellcolor[rgb]{0.91, 0.91, 0.91} 11.83 & \cellcolor[rgb]{0.91, 0.91, 0.91} 11.55 & \cellcolor[rgb]{0.91, 0.91, 0.91} 11.90 & \cellcolor[rgb]{0.91, 0.91, 0.91} 11.41 & \cellcolor[rgb]{0.90, 0.90, 0.90} 12.23 & \cellcolor[rgb]{0.94, 0.94, 0.94} 8.81 & \cellcolor[rgb]{0.98, 0.98, 0.98} 3.61 & \cellcolor[rgb]{0.97, 0.97, 0.97} 4.37 & \cellcolor[rgb]{0.88, 0.88, 0.88} 14.17 & \cellcolor[rgb]{0.95, 0.95, 0.95} 7.74 &\rowid\\   
    &  &  & \algogrpos & \cellcolor[rgb]{0.89, 0.89, 0.89} 13.46 & \cellcolor[rgb]{0.91, 0.91, 0.91} 11.60 & \cellcolor[rgb]{0.92, 0.92, 0.92} 11.16 & \cellcolor[rgb]{0.92, 0.92, 0.92} 11.27 & \cellcolor[rgb]{0.92, 0.92, 0.92} 10.74 & \cellcolor[rgb]{0.91, 0.91, 0.91} 11.65 & \cellcolor[rgb]{0.96, 0.96, 0.96} 5.70 & \cellcolor[rgb]{0.99, 0.99, 0.99} 1.40 & \cellcolor[rgb]{0.98, 0.98, 0.98} 2.91 & \cellcolor[rgb]{0.94, 0.94, 0.94} 8.31 & \cellcolor[rgb]{0.97, 0.97, 0.97} 4.58 &\rowid\\ 
    &  &  & \algogrpo 
      & \cellcolor[rgb]{0.88, 0.88, 0.88} 14.05 & \cellcolor[rgb]{0.91, 0.91, 0.91} 11.73 & \cellcolor[rgb]{0.91, 0.91, 0.91} 11.42 & \cellcolor[rgb]{0.92, 0.92, 0.92} 11.12 & \cellcolor[rgb]{0.92, 0.92, 0.92} 11.37 & \cellcolor[rgb]{0.91, 0.91, 0.91} 11.94 & \cellcolor[rgb]{0.97, 0.97, 0.97} 4.99 & \cellcolor[rgb]{0.99, 0.99, 0.99} 1.57 & \cellcolor[rgb]{0.99, 0.99, 0.99} 2.64 & \cellcolor[rgb]{0.96, 0.96, 0.96} 5.91 & \cellcolor[rgb]{0.98, 0.98, 0.98} 3.78 &\rowid\\
\bottomrule
\end{tabular}
\vspace{-2mm}
    \label{tab:result_1}
\end{table*}

\section{Experiments}

\subsection{Model configuration}
\label{sec:exp:config}
We follow the training pipeline in Figure~\ref{fig:framework} and compare SFT, GRPO, and its variants for fine-tuning (\S~\ref{sec:method}). 

For fine-tuning the post-trained models, we use the post-trained models released by the AntiDeepfake project~\cite{antideepfake_2025}, specifically those using the multi-lingual SSL-based front end: XLS-R-2B~\cite{babu2021xls}, MMS-1B~\cite{mms}, or MMS-300M.\footnote{Corresponding checkpoints are \texttt{xlsr\_2b.ckpt}, \texttt{mms\_1b.ckpt}, and \texttt{mms\_300m.ckpt} (DOI:10.5281/zenodo.15580543).} 
All the models use a similar architecture: the SSL-based front-end processes the input waveform, and the features from its last layer are merged into a fixed-dimensional vector via global average pooling. The vector is then transformed with a linear layer and softmax into the probabilities for $\symboltag{FAKE}$ and $\symboltag{REAL}$. During inference, the logit for $\symboltag{REAL}$ is used for scoring.

For reference, we also include two setups: 1) pre-training plus post-training without fine-tuning and 2) pre-training plus fine-tuning without post-training. In the latter, the front end is initialized using the corresponding SSL model pre-trained by Fairseq~\cite{ott2019fairseq}, and the back end is randomly initialized.

The fine-tuning recipe for \emph{all} the models is the same as AntiDeepfake~\cite{antideepfake_2025} but with changes to fit the computation budget: a maximum number of 10 epochs and early-stop validation per 20k training steps. The checkpoint with the lowest EER on the validation set is used for evaluation. 
For models using GRPO and variants, we follow the DeepSeekMath recipe~\cite{shaoDeepSeekMath2024} and set the number of rollouts and penalty weight to $G=64$ and $\beta=0.04$, respectively. For fine-tuning using SFT and GRPO (SFT$\rightarrow$GRPO), each stage runs at maximum for 10 epochs.

The loop of fine-tuning and evaluation is conducted for three rounds, and the average equal error rate (EER) is reported. All experiments are conducted using a single H100 GPU card. 

\subsection{Data \& protocols}
We use the data protocol from the fine-tuning experiment of AntiDeepfake~\cite{antideepfake_2025}. 
The fine-tuning data (and the target domain) are the training partition of the Deepfake-Eval-2024 dataset (\testdf)~\cite{chandra2025deepfake}.
It contains around 50 hours of real and fake speech data from the social network in 2024, which has a diverse amount of background noise and acoustic conditions. 
To fit the GPU memory, utterances in the fine-tuning set are segmented with a maximum duration of 15 seconds.

The in-domain evaluation data are from the evaluation partition of \testdf. Specifically, we use the five segmented versions provided by~\cite{antideepfake_2025}, which have a maximum utterance duration of 4, 10, 13, 30, and 50 seconds, respectively.
The out-of-domain evaluation datasets include the Audio Deepfake Detection Challenge Track 1.2 evaluation set (\testadd)~\cite{yiADD2023}, FakeOrReal test set (\testfor)~\cite{fakeorreal}, segmented DEEP-VOICE dataset (\testdv)~\cite{bird2023_deepvoice}, and In-the-Wild dataset (\testitw)~\cite{muller22_inthewild}. They cover unseen data with different acoustic conditions from the target domain of \testdf{}, thus are expected to measure the generalization performance of the fine-tuned models.
\footnote{Because the post-trained checkpoints were trained on many databases as well as their evaluation partitions, some commonly used test data (e.g., ASVspoof~\cite{todiscoASVspoof2019Future2019}) cannot be used for evaluation.
}

\subsection{Results comparing GRPO and SFT}
Table~\ref{tab:result_1} lists the average EERs over three rounds of the experiment. 
The first observation is that \textbf{models fine-tuned via SFT show varied degrees of degradation on out-of-domain test sets.}
Compared with the post-trained models without fine-tuning (rows 1, 13, and 17), their SFT fine-tuned counterparts (rows 2, 14, and 18, respectively) achieved lower EERs on the in-domain evaluation sets (\testdf) but higher EERs on most of the out-of-domain test sets. For the post-trained model using XLSR-R-2B, for example, the EER is 1.24\% on \testitw{} (row 1) but increased to 6.35\% after SFT fine-tuning (row 2).  

\textbf{Compared with SFT, fine-tuning using pure \algogrpo{} (or \algogrpos{}) alleviates the degradation on out-of-domain test data while maintaining the performance on in-domain test data.}\footnote{This setup is conceptually similar to the idea of DeepSeek-R0~\cite{guo_deepseek-r1_2025} that conducts RL without SFT. In contrast, fine-tuning via SFT$\rightarrow$GRPO is part of the DeepSeek-R1 pipeline.} With XLS-R-2B, comparison between fine-tuning using SFT (row 2) and \algogrpo{} (row 4) shows that the latter achieved lower EERs on all four out-of-domain test sets. For example, the EER on \testitw{} decreased from 6.35\% (row 2) to 2.19\% (row 4). 
Meanwhile, using \algogrpo{} led to EERs of 10\% on the in-domain test sets, which are on a par with those of using SFT. 
Of course, the EERs on the in-domain test sets are much lower than those without fine-tuning (row 1).

Fine-tuning using \algogrpos{} achieved similar results to the case of using \algogrpo{}, for all three types of detection models (with post-training).  
This indicates that implementation differences, such as clipping reward and keeping a separate `old' parameter set $\hat{\Theta}$, are not critical, at least in our experiment.

From other setups of the experiments, we have two additional observations. First, \textbf{SFT$\rightarrow$GRPO does not outperform GRPO-only fine-tuning}.
While the EERs on the in-domain test sets are comparable,  the former (row 6) obtained higher EERs than the latter (row 4) on the out-of-domain test sets, especially, on \testdv{} (7.04 vs. 2.76\%) and \testitw{} (5.89 vs. 2.19\%). The same trend was observed by comparing SFT$\rightarrow$\algogrpos{} (row 5) and pure \algogrpos{} (row 3). 
Second, \textbf{GRPO is effective when applied to post-trained models, but not to the pre-trained models}. 
This is obvious if we compare the results on the out-of-domain test data in rows 4 and 12. \footnote{In fact, SFT on the post-trained model (row 10) also outperformed SFT on the pre-trained model (row 2) on the out-of-domain test sets, which supports the assumed advantage of post-training~\cite{antideepfake_2025}.
}

In summary, the results indicate that SFT applied to the post-trained model improves the performance on in-domain datasets, but it may lose some knowledge from the post-training and degrade the out-of-domain detection performance. In contrast, the GRPO-tuned model tends to retain the knowledge from post-training while learning from the new training data. 

\subsection{Results of drift analysis}
We further analyzed the drift of data distribution using tools from~\cite{wang_towards_2026}. The overall procedure is to extract the fixed-dimensional embedding vector (after average pooling) for each \emph{fake} utterance in a test set as well as a reference set. An Wasserstein distance, or the drift, between the distribution of the embedding vectors in the test set and that of the reference set are then measured. We followed~\cite{wang_towards_2026} and used the ASVspoof 2019 development set as the reference set. 

We measured the drift using the post-trained XLS-R-2B model, the SFT fine-tuned version, and the GRPO fine-tuned version. The results are plotted in Figure~\ref{fig:result:drift} (a). For the post-trained model (in grey bold profile), the drift measured on the five \testdf{} test sets was higher than the other four test sets (\testadd, \testfor, \testdv, \testitw). This is not surprising because the \testdf{} test sets are relatively newer and more varied in terms of acoustic conditions. SFT fine-tuning using domain-matched data reduced the drift values on the \testdf{} test sets (in  black dotted profile), but the drift on \testdv{} and \testitw{} became higher. 
In contrast, GRPO fine-tuning reduced the drift on the \testdf{} test sets while not making the out-of-domain data drift away. 
This result further supports the finding that fine-tuning using GRPO can alleviate the degradation caused by out-of-domain data.

\subsection{Ablation study on GRPO}
\begin{figure}[!t]
    \centering
    \subfloat[Comparing models using different fine-tuning algorithms]
    {
    \includegraphics[width=\linewidth]{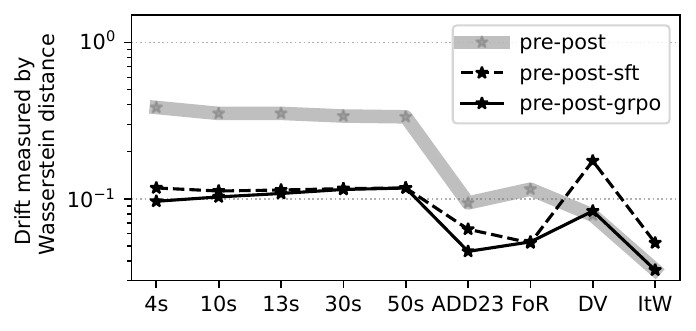}
    \vspace{-2mm}
    }\hfill
    \subfloat[Comparison between different GRPO variants]
    {
    \includegraphics[width=\linewidth]{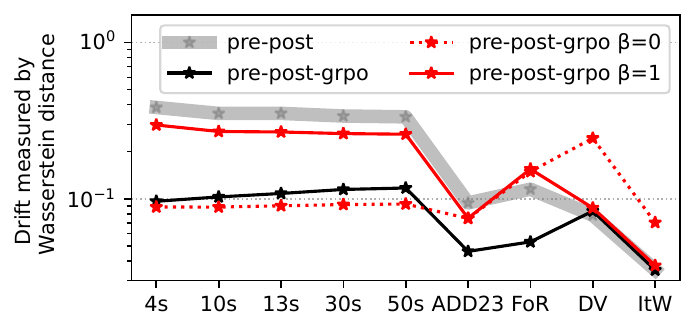}
    \vspace{-2mm}
    }
    \vspace{-2mm}
    \caption{Drift values measured on different test sets}
    \vspace{-2mm}
    \label{fig:result:drift}
\end{figure}

Next, we investigated the impact of the regularization term $\symbolloss{L}_\text{p}$ in GRPO.
Compared with the default setting (row 4), 
\algogrpo{} without regularization ($\beta=0$, row 7) achieved similar EERs on in-domain test sets but an increased EER of 6.50\% on the out-of-domain \testdv{}. 
However, when the regularization was stronger ($\beta=1$, row 8), the model obtained higher EERs than on the in-domain test sets and seemed to under-fit the target domain. 
Hence, unlike studies on LLMs~\cite{yu_dapo_2025,shenfeld_rls_2025}, we cannot conclude that the regularization is nonessential. 
The impact of the regularization term is also supported by the drift analysis in Figure~\ref{fig:result:drift} (b). SFT may also be enhanced with proper regularization, but it is not the focus of our current work.

In terms of the negative reward in GRPO, not using the negative reward (row 9) obtained higher EERs on all the test sets than the default GRPO configuration (row 4). The negative reward may be a key difference between SFT and GRPO. 
While more exploration is needed to explain the improvement by GRPO, it is left to future work.

\section{Conclusion}
In this paper, we investigated whether GRPO-like reinforcement learning is effective in fine-tuning SSL-based speech deepfake detection models. 
Using multiple post-trained detection models with different SSL-based front ends, we conducted experiments and found that, compared with conventional supervised fine-tuning SFT, fine-tuning with GRPO alleviated the degradation on out-of-domain data while maintaining performance on in-domain test sets.

We also identified the preferred recipes of GRPO. Namely, GRPO-only fine-tuning is superior to SFT-then-GRPO; GRPO is effective when it is applied to a post-trained model rather than one using a pre-trained SSL front-end and a randomly initialized back end. 
The negative reward in GRPO was also found to be essential. While the regularization term makes a difference, strong regularization may lead to under-fitting. 
We hope that these findings can inspire more related studies.

\clearpage
\newpage
\section{Acknowledgment}
This work is partially supported by JST, PRESTO Grant (JPMJPR23P9) and a project (JPNP22007) commissioned by the New Energy and Industrial Technology Development Organization (NEDO). The work is done using TSUBAME4.0.

\section{Generative AI Use Disclosure}
Generative AI was ONLY used to fix grammatical errors of the draft. The paper was then human-proofread via a paid service.

\section{References}{
\printbibliography
}

\newpage
\appendix
\onecolumn
\section{Appendix}
\label{sec:app}

\subsection{Notes on the applied GRPO loss}

\subsubsection{GRPO without the negative reward}
In \S~\ref{sec:grpo:setup}, we mentioned that GRPO without any negative sample (\algogrpo{} w/o neg.) is `equivalent' to SFT plus the regularization term. The equivalence is held in terms of the gradient calculated from the GRPO and the SFT loss functions.  To show this, we take the derivative w.r.t $\Theta$ on both sides of Eq.(\ref{eq:rl}) and obtain
\begin{align}
    \nabla_\Theta\symbolloss{L}_{\text{RL}}(\Theta) &= -\symboldomain{E}_{\symbolvec{x}\sim\symbolset{D}, \tilde{y}\sim p_{\widehat{\Theta}}(\symbolvar{y} | \symbolvar{x})}\Big[\symbolloss{A}(\tilde{y},y) \nabla_\Theta \frac{p_\Theta(\tilde{y} | \symbolvec{x})}{p_{\widehat{\Theta}}(\tilde{y} | \symbolvec{x})}-\nabla_\Theta \beta\symbolloss{L}_{\text{p}}({\dot{\Theta}}, {{\Theta}})\Big] \\ 
    &= -\symboldomain{E}_{\symbolvec{x}\sim\symbolset{D}}\Bigg\{\sum_{\tilde{y}}  \Big[\symbolloss{A}(\tilde{y},y)\nabla_\Theta \frac{p_\Theta(\tilde{y} | \symbolvec{x})}{p_{\widehat{\Theta}}(\tilde{y} | \symbolvec{x})}-\nabla_\Theta\beta\symbolloss{L}_{\text{p}}({\dot{\Theta}}, {{\Theta}})\Big] p_{\widehat{\Theta}}(\symbolvar{y} | \symbolvar{x})  \Bigg\} \\
    &= -\symboldomain{E}_{\symbolvec{x}\sim\symbolset{D}}\Bigg\{\sum_{\tilde{y}}  \Big[\symbolloss{A}(\tilde{y},y)\nabla_\Theta p_\Theta(\tilde{y} | \symbolvec{x})-\nabla_\Theta\beta\symbolloss{L}_{\text{p}}({\dot{\Theta}}, {{\Theta}}) p_{\widehat{\Theta}}(\symbolvar{y} | \symbolvar{x})\Big]  \Bigg\} \\
    &= -\symboldomain{E}_{\symbolvec{x}\sim\symbolset{D}}\Bigg\{\sum_{\tilde{y}}  \Big[\symbolloss{A}(\tilde{y},y)\nabla_\Theta p_\Theta(\tilde{y} | \symbolvec{x})\Big] -\nabla_\Theta\beta\symbolloss{L}_{\text{p}}({\dot{\Theta}}, {{\Theta}})  \Bigg\} \\
    &= -\symboldomain{E}_{\symbolvec{x}\sim\symbolset{D}}\Bigg\{\sum_{\tilde{y}}  \Big[\symbolloss{A}(\tilde{y},y) p_\Theta(\tilde{y} | \symbolvec{x}) \nabla_\Theta \log p_\Theta(\tilde{y} | \symbolvec{x})\Big] -\nabla_\Theta\beta\symbolloss{L}_{\text{p}}({\dot{\Theta}}, {{\Theta}})  \Bigg\}.
\end{align}
Note that the last row is based on the Log-Derivative Trick $\nabla_{\theta}f(x;\theta) = f(x;\theta)\nabla_{\theta}\log f(x;\theta) $. If we let $A(\tilde{y}, y)=\delta(\tilde{y} = y)$, i.e., a reward equal to 1 only when the sampled label matches the ground truth, we have 
\begin{align}
    \nabla_\Theta\symbolloss{L}_{\text{RL}}(\Theta) 
    &= -\symboldomain{E}_{\symbolvec{x}\sim\symbolset{D}}\Bigg\{\sum_{\tilde{y}}  \Big[\symbolloss{A}(\tilde{y},y) p_\Theta(\tilde{y} | \symbolvec{x}) \nabla_\Theta \log p_\Theta(\tilde{y} | \symbolvec{x})\Big] -\nabla_\Theta\beta\symbolloss{L}_{\text{p}}({\dot{\Theta}}, {{\Theta}})  \Bigg\}\\
    &= -\symboldomain{E}_{\symbolvec{x}\sim\symbolset{D}}\Bigg\{ p_\Theta({y} | \symbolvec{x}) \nabla_\Theta \log p_\Theta({y} | \symbolvec{x}) -\nabla_\Theta\beta\symbolloss{L}_{\text{p}}({\dot{\Theta}}, {{\Theta}})  \Bigg\}.
\end{align}
Compared with SFT, i.e., $\nabla \symbolloss{L}_{\text{ne}}=-\symboldomain{E}_{\symbolvec{x,y}\sim\symbolset{D}} \Big[\nabla\log{p_\Theta({y} | \symbolvec{x})}\Big]$, the gradient will be similar except the term $p_\Theta(y|\symbolvec{x})$ that weights the gradient $\nabla_\Theta \log p_\Theta(y|\symbolvec{x})$ and regularization term. 

This suggests that GRPO without negative samples prioritizes the model's existing high-probability modes (via the weight $p_\Theta(y|\symbolvec{x})$), behaving like a form of implicit regularization against outlier data. In contrast, the standard SFT treats all (positive) samples equally. This weight term originates from  the ratio $\frac{p_\Theta(\tilde{y} | \symbolvec{x})}{p_{\widehat{\Theta}}(\tilde{y} | \symbolvec{x})}$, i.e., importance sampling ratio used in many RL algorithms.

\subsection{Details on GRPO implementation}
In the implementation, we follow the GRPO paper and added the clipping function: 
\begin{align}
\mathcal{L}_{\text{RL}}(\Theta) = \mathbb{E}_{x \sim \mathcal{D}, \tilde{y}\sim p_{\widehat{\Theta}}(\symbolvar{y} | \symbolvar{x})} \left[ \left( \min \left(  \frac{p_\Theta(\tilde{y}|\symbolvec{x})}{p_{\widehat{\Theta}}(\tilde{y}|\symbolvec{x}) } \mathcal{A}(\tilde{y}, y), \text{clip} \left( \frac{p_\Theta(\tilde{y}|\symbolvec{x})}{p_{\widehat{\Theta}}(\tilde{y}|\symbolvec{x})}, 1-\epsilon, 1+\epsilon \right) \mathcal{A}(\tilde{y}, y) \right) - \beta\symbolloss{L}_{\text{p}}({\dot{\Theta}}, {{\Theta}}) \right) \right],
\end{align}
wherein the advantage function has a floor value in the denominator in case the standard deviation is 0.  \begin{align}
\symbolloss{A}(\tilde{y}_i, \{\tilde{y}_j\}_{j=1}^{G}, y) = \frac{r(\tilde{y}_i, y) - \bar{r}(\{\tilde{y}\}_{j=1}^G, y)}{\sigma(\{\tilde{y}\}_{j=1}^G, y) + \rho},
\end{align}
The hyper-parameters are set to:
\begin{table}[h!]
    \centering
        \caption{Hyper-paramters of the implemented GRPO}
    \begin{tabular}{ll}
    \toprule
    Hyper-parameter & Value \\
    \midrule
        $\epsilon$ & 0.2 \\
        $\rho$ & 1e-5 \\
        \#. of training steps before updating $\widehat{\Theta}$  &  1k \\
        \#. of sampled $\tilde{y}$ per $\symbolvec{x}$ & 64 \\
        \bottomrule
    \end{tabular}

    \label{tab:app:grpo}
\end{table}


\end{document}